

\input harvmac

\input amssym.def
\input amssym
\baselineskip 14pt
\magnification\magstep1
\parskip 6pt

\input epsf

\font \bigbf=cmbx10 scaled \magstep1

\newdimen\itemindent \itemindent=32pt
\def\textindent#1{\parindent=\itemindent\let\par=\resetpar%
\indent\llap{#1\enspace}\ignorespaces}

\let\oldpar=\par
\def\resetpar{\oldpar\parindent=20pt\let\par=\oldpar}

\font\ninerm=cmr9 \font\ninesy=cmsy9
\font\eightrm=cmr8 \font\sixrm=cmr6
\font\eighti=cmmi8 \font\sixi=cmmi6
\font\eightsy=cmsy8 \font\sixsy=cmsy6
\font\eightbf=cmbx8 \font\sixbf=cmbx6
\font\eightit=cmti8
\def\eightpoint{\def\rm{\fam0\eightrm}
  \textfont0=\eightrm \scriptfont0=\sixrm \scriptscriptfont0=\fiverm
  \textfont1=\eighti  \scriptfont1=\sixi  \scriptscriptfont1=\fivei
  \textfont2=\eightsy \scriptfont2=\sixsy \scriptscriptfont2=\fivesy
  \textfont3=\tenex   \scriptfont3=\tenex \scriptscriptfont3=\tenex
  \textfont\itfam=\eightit  \def\it{\fam\itfam\eightit}%
  \textfont\bffam=\eightbf  \scriptfont\bffam=\sixbf
  \scriptscriptfont\bffam=\fivebf  \def\bf{\fam\bffam\eightbf}%
  \normalbaselineskip=9pt
  \setbox\strutbox=\hbox{\vrule height7pt depth2pt width0pt}%
  \let\big=\eightbig  \normalbaselines\rm}
\catcode`@=11 %
\def\eightbig#1{{\hbox{$\textfont0=\ninerm\textfont2=\ninesy
  \left#1\vbox to6.5pt{}\right.\n@@space$}}}
\def\vfootnote#1{\insert\footins\bgroup\eightpoint
  \interlinepenalty=\interfootnotelinepenalty
  \splittopskip=\ht\strutbox %
  \splitmaxdepth=\dp\strutbox %
  \leftskip=0pt \rightskip=0pt \spaceskip=0pt \xspaceskip=0pt
  \textindent{#1}\footstrut\futurelet\next\fo@t}
\catcode`@=12 %

\def\a{\alpha}
\def\b{\beta}
\def\c{\gamma}
\def\d{\delta}

\def\l{\lambda}
\def\m{\mu}
\def\n{\nu}

\def\r{\rho}
\def\s{\sigma}

\def\w{\omega}

\def\C{\Gamma}
\def\D{\Delta}

\def\pl{\partial}

\def\rta{\rightarrow}
\def\lra{\leftrightarrow}

\def\or{\overrightarrow}
\def\Dslash{\,{\raise.15ex\hbox{/}\mkern-12mu D}}

\lref\DH{I.T. Drummond and S. Hathrell, Phys. Rev. D22 (1980) 343. }
\lref\Sch{J.S. Schwinger, Phys. Rev. 82 (1951) 664.}
\lref\DeW{B.S. DeWitt, Dynamical Theory of Groups and Fields (Gordon and Breach,
New York, 1965).}
\lref\Min{S. Minakshisundaram, J. Indian Math. Soc. 17 (1953) 158.}
\lref\See{R.T. Seely, Proc. Symp. Pure Math. 10 (1967) 288.}
\lref\Gilk{P.B. Gilkey, J. Diff. Geom. 10 (1975) 601.}
\lref\Khrip{I.B. Khriplovich, Phys. Lett. B346 (1995) 251.}
\lref\Sthree{G.M. Shore, Nucl. Phys. B460 (1996) 379. }  
\lref\DN{A.D. Dolgov and I.D. Novikov, Phys. Lett. B442 (1998) 82.}
\lref\Gibb{G.W. Gibbons and C.A.R. Herdeiro, Phys. Rev. D63 (2001) 064006.}
\lref\Sfour{G.M. Shore, Nucl. Phys. B605 (2001) 455. }
\lref\Ssix{G.M. Shore, gr-qc/0203034.}
\lref\BGZV{A.O. Barvinsky, Yu.V. Gusev, V.V. Zhytnikov and G.A. Vilkovisky,
Print-93-0274 (Manitoba), 1993.}
\lref\BVone{A.O. Barvinsky and G.A. Vilkovisky, Nucl. Phys. B282 (1987) 163.}
\lref\BVtwo{A.O. Barvinsky and G.A. Vilkovisky, Nucl. Phys. B333 (1990) 471.}
\lref\BVthree{A.O. Barvinsky and G.A. Vilkovisky, Nucl. Phys. B333 (1990) 512.}
\lref\Avram{I.G. Avramidi, Nucl. Phys. B355 (1991) 712.}

\lref\Drum{I.T. Drummond, Phys. Rev. D63 (2001) 043503. } 
\lref\Sone{R.D. Daniels and G.M. Shore, Nucl. Phys. B425 (1994) 634. }
\lref\Stwo{R.D. Daniels and G.M. Shore, Phys. Lett. B367 (1996) 75. }

\lref\LPT{J. I. Latorre, P. Pascual and R. Tarrach, Nucl. Phys. B437
(1995) 60.}
\lref\Gies{W. Dittrich and H. Gies, Phys. Lett. B431 (1998) 420-429;
Phys. Rev. D58 (1998) 025004.}
\lref\Myers{R. Lafrance and  R.C. Myers, Phys. Rev. D51 (1995) 2584.}
\lref\Cho{H.T. Cho, Phys. Rev. D56 (1997) 6416.}
\lref\Cai{R-G. Cai, Nucl. Phys. B524 (1998) 639.}
\lref\Bass{B.A. Bassett, S. Liberati and C. Molina-Paris, Phys. Rev. D62 (2000) 103518.}
\lref\Scharn{K. Scharnhorst, Phys. Lett. B236 (1990) 354.}
\lref\Bart{G. Barton, Phys. Lett. B237 (1990) 559.}
\lref\LSV{S. Liberati, S. Sonego and M. Visser, gr-qc/0107091.}
\lref\LSVtwo{S. Liberati, S. Sonego and M. Visser, Phys. Rev. D63 (2001) 085003.}
\lref\BLV{M. Visser, B. Bassett and S. Liberati, Nucl. Phys. Proc. Suppl. 88
(2000) 267.}

\lref\DNtwo{A.D. Dolgov and I.B. Khriplovich, Phys. Lett. A243 (1998) 117.}
\lref\DK{A.D. Dolgov and I.B. Khriplovich, Sov. Phys. JETP 58(4) (1983) 671.}
         
\lref\Konst{M.Yu. Konstantinov, gr-qc/9810019.}
\lref\Hau{L.V. Hau, Scientific American, July 2001, 66.}
\lref\WKD{L.J. Wang, A. Kuzmich and A. Dogoriu, Nature 406 (2000)  277.}
\lref\Brill{L. Brillouin, {\it Wave Propagation and Group Velocity}, 
Academic Press (London) 1960.}
\lref\Leon{M.A. Leontovich, {\it in} L.I. Mandelshtam,
{\it Lectures in Optics, Relativity and Quantum Mechanics}, Nauka, Moscow 1972~
{(\it in Russian).}}
\lref\HE{S.W. Hawking and G.F.R. Ellis, {\it The Large Scale Structure of
Spacetime}, Cambridge University Press, 1973.}
\lref\Fried{F.G. Friedlander, {\it The Wave Equation on a Curved Spacetime},
Cambridge University Press, 1975.}
\lref\CH{R. Courant and D. Hilbert, {\it Methods of Mathematical Physics, Vol II},
Interscience, New York, 1962.}
\lref\SEF{P. Schneider, J. Ehlers and E.E. Falco, {\it Gravitational Lenses},
Springer-Verlag, New York, 1992.}
\lref\Ch{S. Chandresekhar, {\it The Mathematical Theory of Black Holes},
Clarendon, Oxford, 1985. }
\lref\Inverno{R.A. d'Inverno, {\it Introducing Einstein's Relativity},
Clarendon, Oxford, 1992. }
\lref\Bondi{H. Bondi, M.G.J. van der Burg and A.W.K. Metzner, Proc. Roy. Soc.
A269 (1962) 21.}
\lref\Sachs{R.K. Sachs, Proc. Roy. Soc. A270 (1962) 103. }
\lref\TEone{W. Tsai and T. Erber, Phys. Rev. D10 (1974) 492.}
\lref\TEtwo{W. Tsai and T. Erber, Phys. Rev. D12 (1975) 1132.}
\lref\Adler{S. Adler, Ann. Phys. (N.Y.) 67 (1971) 599.}


{\nopagenumbers
\rightline{SWAT/338}
\vskip1cm
\centerline{\bigbf A Local Effective Action for Photon--Gravity Interactions}

\vskip1cm

\centerline {\bf G.M. Shore}

\vskip0.5cm
\centerline{\it Department of Physics}
\centerline{\it University of Wales, Swansea}
\centerline{\it Singleton Park}
\centerline{\it Swansea, SA2 8PP, U.K.}

\vskip1cm

{
\parindent 1.5cm{

{\narrower\smallskip\parindent 0pt
ABSTRACT:~~Quantum phenomena such as vacuum polarisation in curved spacetime
induce interactions between photons and gravity with quite striking consequences,
including the violation of the strong equivalence principle and the apparent 
prediction of `superluminal' photon propagation. These quantum
interactions can be encoded in an effective action.
In this paper, we extend previous results on the effective action for QED
in curved spacetime due to Barvinsky, Vilkovisky and others and present a
new, local effective action valid to all orders in a derivative expansion, 
as required for a full analysis of the quantum theory of high-frequency 
photon propagation in gravitational fields. 
 
\narrower}}}

\vskip2cm

\leftline{SWAT/338} 
\leftline{May 2002}

\vfill\eject}

\pageno=1

\newsec{Introduction}

In classical electrodynamics in curved spacetime, the interaction of the 
electromagnetic and gravitational fields is encoded in the Maxwell action
\eqn\sectaa{
\C_{(0)} ~~=~~ -{1\over4} \int dx \sqrt{g} ~F_{\m\n}F^{\m\n}
}
where the field strength $F_{\m\n} = D_\m A_\n - D_\n A_\m$ is defined
with covariant derivatives. The interaction involves only the 
connection and is independent of the spacetime curvature. This action 
therefore embodies the strong equivalence principle (SEP), which states 
that the laws of physics should be the same in the local inertial frames 
at each point in spacetime and reduce to their special relativistic form 
at the origin of each of these frames. This is assured since the connections
(though not of course the curvatures) vanish locally in an appropriate frame
in a Riemannian manifold.

The picture is rather different in quantum theory. In quantum electrodynamics
in curved spacetime, vacuum polarisation loops involving the electron 
induce interactions between photons and the background gravitational 
field that depend explicitly on the curvature, with the necessary scale being set
by the Compton wavelength of the electron, $\l_c = {1\over m}$. These interactions 
therefore effectively violate the SEP at the quantum level.

The simplest effective action which encodes some of the essential physics of 
vacuum polarisation was constructed explicitly by Drummond and Hathrell \refs{\DH}, 
who used standard Schwinger--de Witt proper time/heat kernel methods
(see e.g.~\refs{\Sch,\DeW,\Min,\See,\Gilk})
as well as an independent diagrammatic technique to find
\eqnn\sectab
$$\eqalignno{
\C_{\rm DH} ~~=~~ \C_{(0)}  
+ {1\over m^2}\int dx \sqrt{g}~&\biggl(
a R F_{\m\n}F^{\m\n} + b R_{\m\n} F^{\m\l} F^\n{}_\l
+ c R_{\m\n\l\r} F^{\m\n} F^{\l\r} \cr
{}&+ d D_\m F^{\m\l} D_\n F^\n{}_\l ~\biggr) \cr
{}&{}& \sectab \cr }
$$
Here, $a,b,c,d$ are perturbative coefficients of $O(\a)$, viz.
\eqn\sectac{
a = -{1\over144}{\a\over\pi}~~~~~~
b = {13\over360}{\a\over\pi}~~~~~~
c = -{1\over360}{\a\over\pi}~~~~~~
d = -{1\over30}{\a\over\pi}
}
As well as being a one-loop result, there are two other significant
limitations on $\C_{\rm DH}$. First, it is at most of third order in the
generalised `curvatures' (including both the Riemann curvature and the 
electromagnetic field strengths). Second, it is just the lowest-order term
in a derivative expansion involving higher orders in $O({D\over m})$
acting on the curvatures. If we restrict ourselves to the interesting case
of terms of $O(F^2)$, which determine the photon propagator and thus the quantum 
effects on the propagation of light in gravitational fields, the first 
restriction is equivalent to working to first order in $O({R\over m^2})$, or
$O({\l_c^2\over L^2})$ where $L$ is a typical curvature scale, i.e.~weak
background gravitational fields. In terms of photon propagation, the second 
condition implies a restriction to low frequencies, i.e.~lowest order
in $O({\l_c\over \l})$ where $\l$ is the photon wavelength.

By far the most dramatic consequence of eq.\sectab ~is its prediction of 
superluminal photon velocities. As explained in ref.\refs{\DH} (see 
refs.\refs{\Khrip,\Sthree,\DN,\Gibb,\Sfour,\Ssix}
for a selection of subsequent work), the new SEP-violating 
interactions induce curvature-dependent shifts in the light cones, which
become
\eqn\sectad{
k^2 ~-~{1\over m^2}(2b+4c) R_{\m\l} k^\m k^\l ~+~{8c\over m^2} 
C_{\m\n\l\r} k^\m k^\l a^\n a^\r ~~=~~0
}
where $k^\m$ is the photon momentum and $a^\m$ is its polarisation. 
Note that we have written eq.\sectad ~in a convenient form involving the
Weyl tensor $C_{\m\n\l\r}$. For suitably chosen trajectories in certain 
curved spacetimes, this condition remarkably permits $k^2$ to be spacelike.

Although photon propagation is the most striking application of the effective
action, it is clearly important much more generally in encoding the often
unintuitive effects of gravity on quantum fields. In this paper, we 
present an extension of the Drummond-Hathrell action in which we relax
the restriction to lowest order in the derivative expansion, while retaining
the weak gravitational field approximation. The application of this improved 
effective action to the photon propagation problem and the nature of dispersion
in gravitational fields is considered separately in ref.\refs{\Ssix}; here
we discuss the technical issues arising in the construction of the effective
action itself. 

At one-loop order, the QED effective action is given (in Euclidean space)
by
\eqn\sectae{
\C = \C_{(0)} + \ln {\rm det} S(x,x')
}
where $\C_{(0)}$ is the free Maxwell action and $S(x,x')$ is the Green
function of the Dirac operator in the background gravitational field,
i.e.
\eqn\sectaf{
\bigl(i\Dslash  - m\bigr) S(x,x') = {1\over\sqrt{g}} \d(x,x')
}

In fact it is more convenient to work with the differential operator
corresponding to the scalar Green function $G(x,x')$ defined by
\eqn\sectag{
S(x,x') = \bigl(i\Dslash  + m\bigr) G(x,x')
}
so that 
\eqn\sectah{
\Bigl( D^2 + ie\s^{\m\n} F_{\m\n} - {1\over 4} R + m^2\Bigr) G(x,x')~~=~~
-{1\over\sqrt{g}} \d(x,x')
}
Then we evaluate $\C$ from the heat kernel, or proper time, representation
\eqn\sectai{
\C ~~=~~ \C_{(0)} ~-~{1\over2}\int_0^\infty {ds\over s} ~e^{- m^2 s}~ 
{\rm Tr} {\cal G}(x,x';s)
}
where
\eqn\sectaj{
{\cal D}{\cal G}(x,x';s) =  {\pl\over\pl s}{\cal G}(x,x';s)
}
with ${\cal G}(x,x';0) = {1\over\sqrt{g}} \d(x,x')$. 
Here, ${\cal D}$ is the differential operator in eq.\sectah ~at $m=0$.

Our principal result is summarised in the following expression for the
one-loop QED effective action, to first order in $O({R\over m^2})$:
\eqnn\sectak
$$\eqalignno{
\C  = \int dx \sqrt{g} \biggl[-{1\over4}Z~F_{\m\n}F^{\m\n}~
&+~{1\over m^2}\Bigl(D_\m F^{\m\l}~ \or{G_0}~ D_\n F^\n{}_{\l} \cr
&~~~~~~~+~\or{G_1}~ R F_{\m\n} F^{\m\n}~ 
+~\or{G_2}~ R_{\m\n} F^{\m\l}F^\n{}_{\l}~
+~\or{G_3}~ R_{\m\n\l\r}F^{\m\n}F^{\l\r} \Bigr)\cr
&+~{1\over m^4}\Bigl(\or{G_4}~ R D_\m F^{\m\l} D_\n F^\n{}_{\l} \cr
&~~~~~~~+~\or{G_5}~ R_{\m\n} D_\l F^{\l\m}D_\r F^{\r\n}~
+~\or{G_6}~ R_{\m\n} D^\m F^{\l\r}D^\n F_{\l\r} \cr
&~~~~~~~+~\or{G_7}~ R_{\m\n} D^\m D^\n F^{\l\r} F_{\l\r}~
+~\or{G_8}~ R_{\m\n} D^\m D^\l F_{\l\r} F^{\r\n} \cr
&~~~~~~~+~\or{G_9}~ R_{\m\n\l\r} D_\s F^{\s\r}D^\l F^{\m\n}~\Bigr)~~ 
\biggr] \cr 
{}&{}& \sectak \cr }
$$ 
The $\or{G_n}$ ($n\ge 1$) are form factor functions of 
three operators:
\eqn\sectal{
\or{G_n} \equiv G_n\Bigl(-{D_{(1)}^2\over m^2}, -{D_{(2)}^2\over m^2}, 
-{D_{(3)}^2\over m^2}\Bigr)
}
where the first entry ($D_{(1)}^2$) acts on the first following term
(the curvature), etc. $\or{G_0}$ is similarly defined as a single variable 
function. The explicit expressions for these form factors are presented
later in the paper.

Although the entire effective action is a non-local object, for theories
such as QED with a massive electron, it should permit a local expansion
in inverse powers of $m$.
A crucial feature of the above form of the effective action is that it is
indeed manifestly local, in the sense that the form factors 
$\or{G_n}$ have an expansion in positive powers of the $D_{(i)}^2$. 
This depends on making the particular choice of basis operators above. 

A very general effective action calculation, which encompasses the 
more specialised results we need, has been given some time ago 
by Barvinsky, Gusev, Zhytnikov and Vilkovisky (BGZV) in ref.\refs{\BGZV},
building on methods developed earlier by two of the authors \refs{\BVone,
\BVtwo,\BVthree}. Closely related results have also been obtained by
Avramidi \refs{\Avram}. The present paper is essentially a translation and
specialisation of the BGZV result to the $O(RFF)$ terms in the QED effective
action. However, the BGZV result is presented in an apparently non-local
form and the translation into a manifestly local effective action
through an appropriate choice of basis operators involves a number of subtle
manoeuvres. Together with the complexity of the manipulations on the form factors, 
we feel this justifies our independent presentation, in which we explain the 
delicate technical points in the translation and quote explicit results for the 
form factors in the {\it local} effective action \sectak.

In the next section, we introduce the BGZV effective action and show how 
it is recast in local form and reduced to the $O(RFF)$ action we are looking for. 
Then in section 3, we reconsider the $O(F^2)$ terms and rewrite them in a
convenient form consistent with the Drummond-Hathrell action. Section 4 contains 
a summary of our final result, with explicit algebraic expressions for the
form factors collected in an appendix. A numerical analysis of the form factors
is given in section 5. The paper is presented so that the final results for the
local effective action and the form factors are entirely self-contained in 
section 4 and the appendix.

\vskip0.7cm

\newsec{The BGZV Effective Action}

Barvinsky {\it et al.} evaluate the heat kernel  ${\cal G}(x,x';s)$
(denoted by $K$ in refs.\refs{\BGZV,\BVtwo}) corresponding to the generic 
second-order elliptic differential operator
\eqn\sectba{
H = \square \hat{\bf 1} + \hat P - {1\over6}R \hat{\bf 1}, ~~~~~~
\square \equiv g^{\m\n}D_\m D_\n
}
acting on small fluctuations $\d\phi^A$ of an arbitrary set of fields $\phi^A(x)$.
The hat notation denotes matrices acting on the vector space of the $\d\phi^A$, 
i.e. $\hat P = P^A{}_B, ~~~\hat{\bf 1}= \d^A{}_B$, etc.~and the matrix trace is
denoted ${\rm tr}$. The Euclidean metric $g_{\m\n}$ describes the background
spacetime, with gravitational Ricci and Riemann curvatures $R_{\m\n}$ and 
$R_{\m\n\l\r}$ respectively. A further `curvature' is defined by the commutators
of the covariant derivatives acting on the fields $\d\phi^A$:
\eqn\sectbb{
\bigl(D_\m D_\n - D_\n D_\m\bigr)\d\phi^A  = {\cal R}^A{}_{B\m\n}\d\phi^B 
}
and we denote $\hat{\cal R}_{\m\n} = {\cal R}^A{}_{B\m\n}$. The `potential'
$P$ is an arbitrary matrix. BGZV introduce the collective notation ${\frak R}$
for the full set of generalised curvatures:
\eqn\sectbc{
{\frak R} ~~=~~\bigl\{ R_{\m\n\l\r},~\hat{\cal R}_{\m\n},~\hat P \bigr\}
}
and calculate the heat kernel ${\cal G}$ up to $O({\frak R}^3)$.

The first step in extracting the QED effective action from the BGZV results
is to identify the curvatures ${\frak R}$ by comparing the differential
operator ${\cal D}$ of eqs.\sectah,\sectaj ~with the standard form $H$.
In this case, the field $\phi^A$ is the electron spinor field $\psi_\a(x)$,
so we identify $A,B$ as Dirac spinor indices $\a,\b$ and identify
\eqn\sectbd{
\hat P = i e \s^{\m\n} F_{\m\n} - {1\over12}R \hat{\bf 1}
}
The (gauge and gravitational) covariant derivative acting on spinors is
\eqn\sectbe{
D_\m = \pl_\m + {1\over2} \s^{ab}\w_{ab\m} + ieA_\m
}
where $\w^a{}_{b\m}$ is the spin connection
\eqn\sectbf{
\w^a{}_{b\m} = e_b{}^\n \bigl(\pl_\m e^a{}_\n - \C^\l_{\m\n} e^a{}_\l \bigr)
}
with $e^a{}_\m$ the vierbein and $\s^{ab} = {1\over4}[\c^a,\c^b]$.
Evaluating the commutator, we find
\eqn\sectbg{
\bigl[D_\m,D_\n\bigr] = ie F_{\m\n}\hat{\bf 1} + {1\over2} \s^{ab} R_{ab\m\n}
}
and so, comparing with eq.\sectbb, we identify
\eqn\sectbh{
\hat{\cal R}_{\m\n} = ie F_{\m\n} \hat{\bf 1} + {1\over2} \s^{\l\r} R_{\m\n\l\r} 
}

Having established this identification between the BGZV generalised
curvatures and the QED field strength and gravitational curvature tensors,
we can now summarise their result for ${\rm Tr}~{\cal G}(x,x';s)$. Here, 
${\rm Tr}$ denotes the full functional trace, i.e.
${\rm Tr}~{\cal G}(x,x';s) = \int dx \sqrt{g}~
{\rm tr}~{\cal G}(x,x';s)\big|_{x'=x}$.
The result is \refs{\BGZV}:
\eqnn\sectbi
$$\eqalignno{
{\rm Tr}~{\cal G} ~~=~~{1\over (4\pi)^2}{1\over s^2}\int dx \sqrt{g} ~~
{\rm tr}~&\biggl[
\hat{\bf 1} + s \hat P \cr
&+ s^2 \sum_{i=1}^5 f_i(-s\square_{(2)})~ {\frak R}_1 {\frak R}_2 (i) \cr
&+ s^3 \sum_{i=1}^{11} F_i(-s\square_{(1)},-s\square_{(2)},-s\square_{(3)})~ 
{\frak R}_1 {\frak R}_2 {\frak R}_3 (i) \cr
&+ s^4 \sum_{i=12}^{25} F_i(-s\square_{(1)},-s\square_{(2)},-s\square_{(3)})~ 
{\frak R}_1 {\frak R}_2 {\frak R}_3 (i) \cr
&+ s^5 \sum_{i=26}^{28} F_i(-s\square_{(1)},-s\square_{(2)},-s\square_{(3)})~ 
{\frak R}_1 {\frak R}_2 {\frak R}_3 (i) \cr
&+ s^6 ~F_{29}(-s\square_{(1)},-s\square_{(2)},-s\square_{(3)})~ 
{\frak R}_1 {\frak R}_2 {\frak R}_3 (29) ~~+~~O({\frak R}^4) ~~\biggr]\cr
{}&{}& \sectbi \cr }
$$

Here, ${\frak R}_1 {\frak R}_2 (i)$ denote independent basis operators formed
from terms of second order in the generalised curvatures \sectbc. Similarly 
for the independent third order terms ${\frak R}_1 {\frak R}_2 
{\frak R}_3 (i)$. Notice that these appear with different powers of $s$, 
determined by their dimension. The form factors $f_i$ and $F_i$ are functions
of the operators $\square_{(1)}$, etc.~with the suffix indicating which curvature 
term is acted upon. Thus, for example, since ${\frak R}_1 {\frak R}_2 (5) =
\hat{\cal R}_{1\m\n} \hat{\cal R}_2{}^{\m\n}$, the corresponding contribution
to ${\rm Tr}~ {\cal G}$ is
\eqn\sectbj{
{1\over(4\pi)^2} \int dx \sqrt{g}~ \hat{\cal R}_{1\m\n} 
f_5(-s\square) \hat{\cal R}_2{}^{\m\n} 
}

We will return to the detailed expressions for the form factors later. First,
we consider the basis of 5 $O({\frak R}^2)$ and 29 $O({\frak R}^3)$ operators
and, with the identifications in eqs.\sectbd ~and \sectbh, pick out those from 
the full list which produce the terms of $O(F^2)$ and $O(RFF)$ relevant for our 
QED effective action. The remaining terms involving purely gravitational 
curvatures of $O(R^2)$ and $O(R^3)$ are neglected. In BGZV notation, the
relevant basis operators are as follows:
\eqnn\sectbk
$$\eqalignno{
{\rm tr}~{\frak R}_1 {\frak R}_2 (4)~~&=~~ {\rm tr}~ 
\hat P_1 \hat P_2 ~~~~~~~~~~=~~
{1\over2}e^2 F_{\m\n} F^{\m\n} ~{\rm tr}\hat{\bf 1} ~~+~~O(R^2) \cr
{\rm tr}~{\frak R}_1 {\frak R}_2 (5)~~&=~~ {\rm tr}~
\hat{\cal R}_{1\m\n} \hat{\cal R}_2{}^{\m\n} ~~~=~~
- e^2 F_{\m\n} F^{\m\n} ~{\rm tr}\hat{\bf 1} ~~+~~O(R^2) \cr 
{\rm tr}~{\frak R}_1 {\frak R}_2 {\frak R}_3 (1)  ~~&=~~ {\rm tr}~
\hat P_1 \hat P_2 \hat P_3 ~~=~~
-{1\over24} e^2\Bigl(RF_{\m\n}F^{\m\n} + F_{\m\n}RF^{\m\n} + F_{\m\n}F^{\m\n}R
\Bigr)  ~{\rm tr}\hat{\bf 1} ~+~O(R^3) \cr
{\rm tr}~{\frak R}_1 {\frak R}_2 {\frak R}_3 (3)  ~~&=~~ {\rm tr}~
\hat{\cal R}_{1\m\n} \hat{\cal R}_2^{\m\n} \hat P_3 ~~=~
\Bigl({1\over12}e^2F_{\m\n}F^{\m\n}R + 
{1\over4}e^2 R_{\m\n\l\r}F^{\m\n}F^{\l\r} \cr
&~~~~~~~~~~~~~~~~~~~~~~~~~~~~~~~~~~~~~~~~~~~~~~~~~~~~~~~~
+ {1\over4}e^2 F^{\m\n}R_{\m\n\l\r}F^{\l\r} \Bigr)~
{\rm tr}\hat{\bf 1} ~+~O(R^3) \cr
{\rm tr}~{\frak R}_1 {\frak R}_2 {\frak R}_3 (6)  ~~&=~~ {\rm tr}~
\hat P_1 \hat P_2 R_3 ~~=~
{1\over2}e^2 F_{\m\n}F^{\m\n} R ~{\rm tr}\hat{\bf 1} ~+~O(R^3) \cr
{\rm tr}~{\frak R}_1 {\frak R}_2 {\frak R}_3 (7)  ~~&=~~ {\rm tr}~
R_1 \hat{\cal R}_{2\m\n} \hat{\cal R}_3^{\m\n}~~=~
-e^2 R F_{\m\n}F^{\m\n} ~{\rm tr}\hat{\bf 1} ~+~O(R^3) \cr
{\rm tr}~{\frak R}_1 {\frak R}_2 {\frak R}_3 (8)  ~~&=~~ {\rm tr}~
R_{1\m\n} \hat{\cal R}_2^{\m\l} \hat{\cal R}_3^\n{}_{\l}~~=~
-e^2 R_{\m\n} F^{\m\l} F^\n{}_\l ~{\rm tr}\hat{\bf 1} ~+~O(R^3) \cr
{\rm tr}~{\frak R}_1 {\frak R}_2 {\frak R}_3 (14)  ~~&=~~ {\rm tr}~
D_\m \hat{\cal R}_1^{\m\l} D_\n \hat{\cal R}_2^\n{}_{\l} \hat P_3 ~~=~
\Bigl({1\over12}e^2 D_\m F^{\m\l} D_\n F^\n{}_\l R
+{1\over4}e^2 D_\m R_{\r\s}{}^{\m\l} D_\n F^\n{}_\l F^{\r\s} \cr
&~~~~~~~~~~~~~~~~~~~~~~~~~~~~~~~~~~~~~~~~~~~~~~~~
+{1\over4}e^2 D_\m F^{\m\l} D_\n R_{\r\s}{}^\n{}_\l F^{\r\s}\Bigr)~
{\rm tr}\hat{\bf 1} ~+~O(R^3) \cr
{\rm tr}~{\frak R}_1 {\frak R}_2 {\frak R}_3 (17)  ~~&=~~ {\rm tr}~
R_{1\m\n} D^\m D^\n \hat P_2 \hat P_3 ~~=~
{1\over2}e^2 R_{\m\n} D^\m D^\n F^{\l\r} F_{\l\r}~
{\rm tr}\hat{\bf 1} ~+~O(R^3) \cr
{\rm tr}~{\frak R}_1 {\frak R}_2 {\frak R}_3 (18)  ~~&=~~ {\rm tr}~
R_{1\m\n} D_\l \hat{\cal R}_2^{\l\m} D_\r \hat{\cal R}_3^{\r\n}~~=~
-e^2 R_{\m\n} D_\l F^{\l\m} D_\r F^{\r\n}~
{\rm tr}\hat{\bf 1} ~+~O(R^3) \cr
{\rm tr}~{\frak R}_1 {\frak R}_2 {\frak R}_3 (19)  ~~&=~~ {\rm tr}~
R_{1\m\n} D^\m \hat{\cal R}_2^{\l\r} D^\n \hat{\cal R}_{3\l\r}~~=~
-e^2 R_{\m\n}  D^\m F^{\l\r} D^\n F_{\l\r}~
{\rm tr}\hat{\bf 1} ~+~O(R^3) \cr
{\rm tr}~{\frak R}_1 {\frak R}_2 {\frak R}_3 (20)  ~~&=~~ {\rm tr}~
R_1 D_\m \hat{\cal R}_2^{\m\l} D_\n \hat{\cal R}_3{}^\n{}_\l~~=~
-e^2 R D_\m F^{\m\l} D_\n F^\n{}_\l ~
{\rm tr}\hat{\bf 1} ~+~O(R^3) \cr
{\rm tr}~{\frak R}_1 {\frak R}_2 {\frak R}_3 (21)  ~~&=~~ {\rm tr}~
R_{1\m\n} D^\m D^\l \hat{\cal R}_{2\l\r}  \hat{\cal R}_3^{\r\n}~~=~
-e^2 R_{\m\n} D^\m D^\l F_{\l\r} F^{\r\n}~
{\rm tr}\hat{\bf 1} ~+~O(R^3) \cr
{}&{}& \sectbk \cr }
$$
Notice that for clarity we have omitted the subscripts 1,2,3 on the terms 
in the right-hand column, but must remember that the order is still significant
since they are to be acted on by the form factors. $\hat{\bf 1}$ is the 
unit Dirac matrix, so in four dimensions ${\rm tr}\hat{\bf 1} = 4$.

Collecting these terms and substituting into eq.\sectbi ~then gives the 
complete $O(F^2)$ and $O(RFF)$ terms in the BGZV effective action.
However, as it stands this is expressed in {\it non-local} form, whereas we
require a manifestly local action. As explained in ref.\refs{\BVtwo}, the key 
is to reintroduce the Riemann tensor into the basis of operators. (Notice
that none of the above set of terms in the form ${\rm tr}~{\frak R}_1 
{\frak R}_2 {\frak R}_3$ involves the uncontracted Riemann tensor $R_{\m\n\l\r}$,
which only enters in the right-hand column when we expand $\hat{\cal R}_{\m\n}$
and $\hat P$ according to eqs.\sectbd,\sectbh. This is a feature of the
covariant perturbation theory construction developed by Barvinsky and
Vilkovisky.) We therefore introduce the further operator
\eqn\sectbl{
{\rm tr}~R_{1\m\n\l\r} \hat{\cal R}_2^{\m\n} \hat{\cal R}_3^{\l\r} ~~=~~
-e^2 R_{\m\n\l\r} F^{\m\n} F^{\l\r}~{\rm tr} \hat{\bf 1} ~~+~~O(R^3)
}

The Riemann tensor admits the following non-local re-expression in terms of the
Ricci tensor, up to $O(R^2)$:
\eqn\sectbm{
R_{\m\n\l\r} ~~=~~{1\over\square} \Big(D_\m D_\l R_{\n\r} - D_\n D_\l R_{\m\r}
- D_\m D_\r R_{\n\l} + D_\n D_\r R_{\m\l} \Bigr) ~+~O(R^2)
}
This is proved \refs{\BVtwo} by differentiating the Bianchi identity
\eqn\sectbn{
R_{\m\n[\l\r;\s]} ~~=~~0
}
to obtain an equation for $\square R_{\m\n\l\r}$, which is solved iteratively
in powers of curvature using the Green function $1/\square$ to obtain \sectbm.

We can then rewrite the four BGZV basis operators which turn out to have non-local
form factors as:
\eqnn\sectbo
$$\eqalignno{
\int dx \sqrt{g}~{\rm tr}~&\biggl[ 
s F_8 ~R_{\m\n} \hat{\cal R}^{\m\l}  \hat{\cal R}^\n{}_{\l}~~+~~
s^2 F_{18} ~R_{\m\n} D_\l \hat{\cal R}^{\l\m} D_\r \hat{\cal R}^{\r\n} \cr
&+~s^2 F_{19} ~R_{\m\n} D^\m \hat{\cal R}^{\l\r} D^\n \hat{\cal R}_{\l\r}~~+~~
s^2 F_{21} ~R_{\m\n} D^\m D^\l \hat{\cal R}_{\l\r} \hat{\cal R}^{\r\n} 
\biggr] \cr
=~~ \int dx \sqrt{g}~{\rm tr}~&\biggl[
s \tilde F_8 ~R_{\m\n} \hat{\cal R}^{\m\l}  \hat{\cal R}^\n{}_{\l}~~+~~
s^2 \tilde F_{18} ~R_{\m\n} D_\l \hat{\cal R}^{\l\m} D_\r \hat{\cal R}^{\r\n} \cr
&+~s^2\tilde F_{19} ~R_{\m\n} 
D^\m \hat{\cal R}^{\l\r} D^\n \hat{\cal R}_{\l\r}~~+~~
s^2 \tilde F_{21} ~R_{\m\n} D^\m D^\l \hat{\cal R}_{\l\r} \hat{\cal R}^{\r\n} \cr
&+~s\tilde F_0~R_{\m\n\l\r} \hat{\cal R}^{\m\n} \hat{\cal R}^{\l\r} \biggr] \cr
{}&{}& \sectbo \cr }
$$
with the following relation between the form factors:
\eqnn\sectbp
$$\eqalignno{
F_8(x_1,x_2,x_3)~~ &=~~ \tilde F_8(x_1,x_2,x_3)~~
+~~ 2\Bigl(1+{x_2 + x_3\over x_1}\Bigr) \tilde F_0(x_1,x_2,x_3) \cr
F_{18}(x_1,x_2,x_3) ~~&=~~ \tilde F_{18}(x_1,x_2,x_3)~~
-~~ {4\over x_1} \tilde F_0(x_1,x_2,x_3) \cr
F_{19}(x_1,x_2,x_3) ~~&=~~ \tilde F_{19}(x_1,x_2,x_3)~~
+~~ {2\over x_1} \tilde F_0(x_1,x_2,x_3) \cr
F_{21}(x_1,x_2,x_3) ~~&=~~ \tilde F_{21}(x_1,x_2,x_3)~~
-~~ {8\over x_1} \tilde F_0(x_1,x_2,x_3) \cr
{}&{}& \sectbp \cr }
$$ 
where we have used the notation $x_i = -\square_{(i)}$. (Notice that
BGZV \refs{\BGZV} use $\xi_i = -s \square_{(i)}$ in discussing the form
factors.)

The new form factor $\tilde F_0(x_1,x_2,x_3)$ is defined so as to remove
the singular terms in $F_{8,18,19,21}$ leaving the $\tilde F_{8,18,19,21}$
form factors regular as $x_1, x_2, x_3 \rta 0$. Of course this under-determines 
$\tilde F_0$, but any ambiguity only amounts to a reshuffling of contributions
amongst the various terms in the final expression for the effective action.

In the local basis, we therefore find that all the $O(F^2)$ and $O(RFF)$ terms
in the heat kernel are contained in the following BGZV terms (again
suppressing the arguments of the form factors, i.e. $f_i \equiv f_i(-s 
\square_{(2)})$ and 
$F_i \equiv F_i(-s \square_{(1)},-s \square_{(2)},-s \square_{(3)})$):
\eqnn\sectbq
$$\eqalignno{
\int dx \sqrt{g}~{\rm tr}~&\biggl[~
f_4~ \hat P_1 \hat P_2 ~~+~~f_5~ \hat{\cal R}_{1\m\n} \hat{\cal R}_2{}^{\m\n} \cr
&+~ sF_1~\hat P_1 \hat P_2 \hat P_3 ~~
+~~ sF_3~\hat{\cal R}_{1\m\n} \hat{\cal R}_2^{\m\n} \hat P_3 ~~
+~~sF_6~\hat P_1 \hat P_2 R_3~~
+~~sF_7~R_1 \hat{\cal R}_{2\m\n} \hat{\cal R}_3^{\m\n}\cr
&+~s^2 F_{14}~D_\m \hat{\cal R}_1^{\m\l} D_\n \hat{\cal R}_2^\n{}_{\l} \hat P_3~~
+~~s^2 F_{17}~R_{1\m\n} D^\m D^\n \hat P_2 \hat P_3~~
+~~s^2 F_{20}~R_1 D_\m \hat{\cal R}_2^{\m\l} D_\n \hat{\cal R}_3{}^\n{}_\l \cr
&+~s \tilde F_8~R_{1\m\n} \hat{\cal R}_2^{\m\l} \hat{\cal R}_3^\n{}_{\l}~~
+~~s^2 \tilde F_{18}~R_{1\m\n} D_\l \hat{\cal R}_2^{\l\m} D_\r \hat{\cal R}_3^{\r\n}\cr
&+~s^2 \tilde F_{19}~R_{1\m\n} D^\m \hat{\cal R}_2^{\l\r} D^\n \hat{\cal R}_{3\l\r}~~
+~~s^2 \tilde F_{21}~R_{1\m\n} D^\m D^\l \hat{\cal R}_{2\l\r}  \hat{\cal R}_3^{\r\n}\cr
&+~s \tilde F_0~ R_{\m\n\l\r} \hat{\cal R}^{\m\n} \hat{\cal R}^{\l\r} ~~\biggr] \cr
{}&{}& \sectbq \cr }
$$

We can now re-express this explicitly in terms of the $O(F^2)$ and $O(RFF)$ basis 
operators of eq.\sectak ~using the set of relations \sectbk.
We also need to use the following identity, which is proved by integrating by
parts and using the Bianchi identity for $F_{\m\n}$:
\eqnn\sectbr
$$\eqalignno{
&\int dx \sqrt{g} ~D^\l R_{\m\n\l\r} D_\s F^{\s\r} F^{\m\n} \cr
&=~~-~\int dx \sqrt{g}~\biggl(~R_{\m\n\l\r}D_\s F^{\s\r}D^\l F^{\m\n} ~~
+~~{1\over2} R_{\m\n\l\r}\square F^{\m\n} F^{\l\r} ~~+~~O(R^2) ~\biggr) \cr
{}&{}& \sectbr \cr }
$$
This gives:
\eqnn\sectbs
$$\eqalignno{
{\rm Tr}~{\cal G}~~=~~-{e^2\over(4\pi)^2} {\rm tr}\hat{\bf 1}~ \int dx \sqrt{g}~ 
&\biggl[~~ F_{\m\n}~ h_0~F^{\m\n} \cr
&+~s~\Bigl( h_1~R F_{\m\n} F^{\m\n}~ 
+~ h_2~ R_{\m\n} F^{\m\l}F^\n{}_{\l}~
+~ h_3~ R_{\m\n\l\r}F^{\m\n}F^{\l\r} \Bigr)\cr
&+~s^2~\Bigl( h_4~R D_\m F^{\m\l} D_\n F^\n{}_{\l} \cr
&{}~~~~~+~h_5~R_{\m\n} D_\l F^{\l\m}D_\r F^{\r\n}~
+~h_6~R_{\m\n} D^\m F^{\l\r}D^\n F_{\l\r} \cr
&{}~~~~~+~h_7~R_{\m\n} D^\m D^\n F^{\l\r} F_{\l\r}~
+~h_8~R_{\m\n} D^\m D^\l F_{\l\r} F^{\r\n} \cr
&{}~~~~~+~h_9~R_{\m\n\l\r} D_\s F^{\s\r}D^\l F^{\m\n}~\Bigr)~~ 
\biggr] \cr
{}&{}& \sectbs \cr }
$$
where $h_0 \equiv h_0(-s \square )$ and
$h_i \equiv h_i(-s \square_{(1)},-s \square_{(2)},-s \square_{(3)}),~~i\ge 1$.
The form factors are related to the BGZV definitions as follows 
(where we assume appropriate symmetrisations on $x_1,x_2, x_3$):
\eqnn\sectbt
$$\eqalignno{
h_0(x) ~~&=~~-{1\over2} f_4(x) ~~+~~f_5(x) \cr
h_1(x_1,x_2,x_3) ~~&=~~{1\over8}F_1(x_1,x_2,x_3) ~-~
{1\over12}F_3(x_2,x_3,x_1) \cr
&~~~~~~-{1\over2}F_6(x_2,x_3,x_1) ~+~
F_7(x_1,x_2,x_3) \cr
h_2(x_1,x_2,x_3) ~~&=~~\tilde F_8(x_1,x_2,x_3) \cr
h_3(x_1,x_2,x_3) ~~&=~~-{1\over2} F_3(x_1,x_2,x_3) ~-~
{1\over8} F_{14}(x_1,x_2,x_3) (x_2 + x_3) \cr
&~~~~~~+~ \tilde F_0(x_1,x_2,x_3) \cr
h_4(x_1,x_2,x_3) ~~&=~~-{1\over12} F_{14}(x_2,x_3,x_1) ~+~
F_{20}(x_1,x_2,x_3)\cr
h_5(x_1,x_2,x_3) ~~&=~~\tilde F_{18}(x_1,x_2,x_3) \cr
h_6(x_1,x_2,x_3) ~~&=~~\tilde F_{19}(x_1,x_2,x_3) \cr
h_7(x_1,x_2,x_3) ~~&=~~-{1\over2} F_{17}(x_1,x_2,x_3) \cr
h_8(x_1,x_2,x_3) ~~&=~~\tilde F_{21}(x_1,x_2,x_3) \cr
h_9(x_1,x_2,x_3) ~~&=~~{1\over2} F_{14}(x_1,x_2,x_3) \cr
{}&{}& \sectbt \cr }
$$

\vskip0.7cm

\newsec{Second Order Terms}

While eq.\sectbs ~is a perfectly satisfactory, local expression for 
${\rm Tr}~{\cal G}$, it is not the most convenient form for many applications, 
e.g.~photon propagation.
Nor is it the form which permits the simplest comparison with standard
Schwinger--deWitt results. The key step in transforming to this standard form
is to rewrite the second-order contributions of the type $\int dx \sqrt{g}~
F_{\m\n}\square F^{\m\n}$ as $\int dx \sqrt{g}~D_\m F^{\m\l} D_\n F^\n{}_\l$,
plus higher order terms of $O(RFF)$ which of course modify the corresponding
form factors. This second choice for the $O(F^2)$ basis operator also matches
the term in the Drummond-Hathrell action \sectab ~and turns out to be the
more useful representation. 

This transformation can be made already at the level of the BGZV curvatures using
the identity \refs{\BVtwo,\BGZV}:
\eqnn\sectca
$$\eqalignno{
{\rm tr}~\int dx \sqrt{g}~ \hat{\cal R}_{\m\n} \square \hat{\cal R}^{\m\n}~~=~~
{\rm tr}~\int dx \sqrt{g}~\Bigl(
&-2 D_\m \hat{\cal R}^{\m\l} D_\n \hat{\cal R}^\n{}_\l ~+~
4 \hat{\cal R}^\m{}_\l \hat{\cal R}^\l{}_\r \hat{\cal R}^\r{}_\m \cr
&+2R_{\m\n} \hat{\cal R}^{\m\l} \hat{\cal R}^\n{}_\l ~-~
R_{\m\n\l\r} \hat{\cal R}^{\m\n} \hat{\cal R}^{\l\r} \Bigr) \cr
{}&{}& \sectca \cr }
$$
In terms of the electromagnetic field strength and gravitational curvature,
this is equivalent to
\eqnn\sectcb
$$\eqalignno{
\int dx \sqrt{g}~F_{\m\n} \square F^{\m\n} ~~=~~
\int dx \sqrt{g}~\Bigl(&-2 D_\m F^{\m\l} D_\n F^\n{}_\l  \cr
&+~2 R_{\m\n} F^{\m\l} F^\n{}_\l ~-~ R_{\m\n\l\r} F^{\m\n} F^{\l\r} ~+~
O(R^2)~\Bigr) \cr
{}&{}& \sectcb \cr }
$$
The proof is straightforward, relying on the Bianchi identity for $F_{\m\n}$
and the cyclic symmetry of the Riemann tensor.

We now need to extend this identity to include the full form factors.
The first step is to separate off the constant term in $h_0(-\square)$ and
define
\eqn\sectcc{
\int dx \sqrt{g}~F_{\m\n} h_0(-\square) F^{\m\n} ~~=~~
h_0(0)~\int dx \sqrt{g}~F_{\m\n} F^{\m\n} ~+~
\int dx \sqrt{g}~F_{\m\n} h(-\square) \square F^{\m\n}
}
The first term on the r.h.s., where $h(0) = -1/6$, eventually gives rise
to a divergent contribution to the effective action which is removed by the 
conventional electromagnetic field renormalisation. The remaining term simplifies
in two stages. First, write
\eqnn\sectcd
$$\eqalignno{
\int dx \sqrt{g}~F_{\m\n}  h(-\square) \square F^{\m\n} ~~=~~
\int dx \sqrt{g}~\Bigl(&-2 D_\m F^{\m\l}  h(-\square) D_\n F^\n{}_\l ~+~
2 R_{\m\n} F^{\m\l} h(-\square) F^\n{}_\l \cr
&- R_{\m\n\l\r} F^{\m\n}  h(-\square) F^{\l\r}~+~
2 D_\m F^{\m\l} \bigl[ D_\n, h(-\square)\bigr] F^\n{}_\l ~\Bigr) \cr
{}&{}& \sectcd \cr }
$$
The second and third terms are already in the required form of $O(RFF)$ basis
operators. Neglecting higher-order terms in the curvature simplifies the
commutator, and with repeated use of the Bianchi identity for the Riemann
tensor we eventually find
\eqnn\sectce
$$\eqalignno{
\int dx \sqrt{g}~F_{\m\n}  h(-\square) \square F^{\m\n} ~~=~~
&\int dx \sqrt{g}~\biggl(-2 D_\m F^{\m\l}  h(-\square) D_\n F^\n{}_\l \cr
&+2 R_{\m\n} \bigl( h(-\square)+ \square h'(-\square)\bigr) 
F^{\m\l} F^\n{}\l ~-~
R_{\m\n\l\r} h(-\square)F^{\m\n} F^{\l\r} \cr
&-2 R_{\m\n} h'(-\square)D_\l F^{\l\m} D_\r F^{\r\n} ~+~
4 R_{\m\n} h'(-\square)D^\m D^\l F_{\l\r} F^{\r\n} \cr
&+2 R_{\m\n\l\r} h'(-\square)D_\s F^{\s\r} D^\l F^{\m\n} ~~+~
O(R^2 FF)~\biggr) \cr
{}&{}& \sectce \cr }
$$
The transformation between the alternative choices for the $O(F^2)$ basis
operator therefore affects the $O(RFF)$ form factors
$h_2, h_3, h_5, h_8$ and $h_9$ , in the notation of \sectbs.

With this transformation, we can rewrite ${\rm Tr}~{\cal G}$ into the preferred 
form:
\eqnn\sectcf
$$\eqalignno{
{\rm Tr}~{\cal G}~~=~~-{e^2\over(4\pi)^2} {\rm tr}\hat{\bf 1}~ \int dx \sqrt{g}~ 
&\biggl[~~ -{1\over6}F_{\m\n}F^{\m\n}  + s~D_\m F^{\m\l} ~g_0~ D_\n F^\n{}_\l \cr
&+~s~\Bigl( g_1~R F_{\m\n} F^{\m\n}~ 
+~ g_2~ R_{\m\n} F^{\m\l}F^\n{}_{\l}~
+~ g_3~ R_{\m\n\l\r}F^{\m\n}F^{\l\r} \Bigr)\cr
&+~s^2~\Bigl( g_4~R D_\m F^{\m\l} D_\n F^\n{}_{\l} \cr
&{}~~~~~+~g_5~R_{\m\n} D_\l F^{\l\m}D_\r F^{\r\n}~
+~g_6~R_{\m\n} D^\m F^{\l\r}D^\n F_{\l\r} \cr
&{}~~~~~+~g_7~R_{\m\n} D^\m D^\n F^{\l\r} F_{\l\r}~
+~g_8~R_{\m\n} D^\m D^\l F_{\l\r} F^{\r\n} \cr
&{}~~~~~+~g_9~R_{\m\n\l\r} D_\s F^{\s\r}D^\l F^{\m\n}~\Bigr)~~ 
\biggr]  \cr
{}&{}& \sectcf \cr }
$$
with $g_0 \equiv g_0(-s \square )$ and
$g_i \equiv g_i(-s \square_{(1)},-s \square_{(2)},-s \square_{(3)}),~~i\ge 1$
as before. The new form factors are related to the initial ones by:
\eqnn\sectcg
$$\eqalignno{
g_0(x) ~~&=~~-2h(x) ~~=~~
{2\over x} \Bigl(-{1\over2} f_4(x) ~+~f_5(x)~+~{1\over6}\Bigr) \cr
g_{1,4,6,7} ~~&=~~h_{1,4,6,7} \cr
g_2 ~~&=~~h_2 + h(x_2) +x_2 h'(x_2) + h(x_3) +x_3 h'(x_3)\cr
g_3 ~~&=~~h_3 -{1\over2}\bigl(h(x_2) + h(x_3)\bigr) \cr
g_5 ~~&=~~h_5 -\bigl(h'(x_2) + h'(x_3) \bigr) \cr
g_8 ~~&=~~h_8 - 4 h'(x_2) \cr
g_9 ~~&=~~h_9 + 2 h'(x_2) \cr
{}&{}& \sectcg \cr }
$$
where $g_i \equiv g_i(x_1,x_2,x_3), ~~i\ge 1$ and 
we have symmetrised where appropriate.

\vfill\eject

\newsec{The Local Effective Action}

Collecting these results and evaluating the effective action from
\eqn\sectda{
\C ~~=~~ \C_0 ~-~ {1\over2}\int_0^\infty {ds\over s}~e^{-m^2 s}~
{\rm Tr}~{\cal G}(x,x';s)
}
we find the renormalised expression:
\eqnn\sectdb
$$\eqalignno{
\C  = \int dx \sqrt{g} \biggl[-{1\over4}Z~F_{\m\n}F^{\m\n}~
&+~{1\over m^2}\Bigl(D_\m F^{\m\l}~ \or{G_0}~ D_\n F^\n{}_{\l} \cr
&~~~~~~~+~\or{G_1}~ R F_{\m\n} F^{\m\n}~ 
+~\or{G_2}~ R_{\m\n} F^{\m\l}F^\n{}_{\l}~
+~\or{G_3}~ R_{\m\n\l\r}F^{\m\n}F^{\l\r} \Bigr)\cr
&+~{1\over m^4}\Bigl(\or{G_4}~ R D_\m F^{\m\l} D_\n F^\n{}_{\l} \cr
&~~~~~~~+~\or{G_5}~ R_{\m\n} D_\l F^{\l\m}D_\r F^{\r\n}~
+~\or{G_6}~ R_{\m\n} D^\m F^{\l\r}D^\n F_{\l\r} \cr
&~~~~~~~+~\or{G_7}~ R_{\m\n} D^\m D^\n F^{\l\r} F_{\l\r}~
+~\or{G_8}~ R_{\m\n} D^\m D^\l F_{\l\r} F^{\r\n} \cr
&~~~~~~~+~\or{G_9}~ R_{\m\n\l\r} D_\s F^{\s\r}D^\l F^{\m\n}~\Bigr)~~ 
\biggr] \cr 
{}&{}& \sectdb \cr }
$$ 
where the form factors $\or{G_n} = G_n\bigl(-{\square_{(1)}\over m^2},
-{\square_{(2)}\over m^2},-{\square_{(3)}\over m^2}\bigr)$ are defined as
proper time integrals of the $g_n$, viz.
\eqn\sectdc{
G_n(x_1,x_2,x_3) ~~=~~ -{1\over2} {\a\over\pi} ~\int _0^\infty {ds\over s}~
e^{-s} s^p~g_n(sx_1,sx_2,sx_3)
}
where $p=1$ for $n=0,\ldots 3$  and $p=2$ for $n=4,\ldots 9$.
Note that we have rescaled $s$ here so that it is dimensionless, while now
$x_i = -\square_{(i)}/m^2$. The divergent contribution to the 
$F_{\m\n}F^{\m\n}$ term has been evaluated using dimensional regularisation
and cancelled by the standard electromagnetic field renormalisation. The
residual finite $O(\a)$ contribution gives
$Z = 1 + {1\over6} {\a\over\pi}\ln{m^2\over\bar\m^2}$ for a suitable choice of 
RG scale $\bar \m$.

The form factors are given in terms of the BGZV definitions by the following set
of relations:
\eqnn\sectdd
$$\eqalignno{
g_0(x) ~~&=~~-2h(x)~~=~~
{2\over x} \Bigl(-{1\over2} f_4(x) ~+~f_5(x)~+~{1\over6}\Bigr) \cr
g_1(x_1,x_2,x_3) ~~&=~~{1\over8}F_1(x_1,x_2,x_3) ~-~
{1\over12}F_3(x_2,x_3,x_1) \cr
&~~~~~~-{1\over2}F_6(x_2,x_3,x_1) ~+~ F_7(x_1,x_2,x_3) \cr
g_2(x_1,x_2,x_3) ~~&=~~\tilde F_8(x_1,x_2,x_3) ~+~ 
h(x_2) +x_2 h'(x_2) + h(x_3) +x_3 h'(x_3)\cr
g_3(x_1,x_2,x_3) ~~&=~~-{1\over2} F_3(x_1,x_2,x_3) ~-~
{1\over8} F_{14}(x_1,x_2,x_3) (x_2 + x_3) \cr
&~~~~~~+~ \tilde F_0(x_1,x_2,x_3) ~-~{1\over2}\bigl(h(x_2) + h(x_3)\bigr)\cr
g_4(x_1,x_2,x_3) ~~&=~~-{1\over12} F_{14}(x_2,x_3,x_1) ~+~
F_{20}(x_1,x_2,x_3)\cr
g_5(x_1,x_2,x_3) ~~&=~~\tilde F_{18}(x_1,x_2,x_3) ~-~
\bigl(h'(x_2) + h'(x_3) \bigr)\cr
g_6(x_1,x_2,x_3) ~~&=~~\tilde F_{19}(x_1,x_2,x_3) \cr
g_7(x_1,x_2,x_3) ~~&=~~-{1\over2} F_{17}(x_1,x_2,x_3) \cr
g_8(x_1,x_2,x_3) ~~&=~~\tilde F_{21}(x_1,x_2,x_3) ~-~ 4 h'(x_2)\cr
g_9(x_1,x_2,x_3) ~~&=~~{1\over2} F_{14}(x_1,x_2,x_3)~+~ 2 h'(x_2) \cr
{}&{}& \sectdd \cr }
$$
where $h'(x) \equiv {dh\over dx}$ and the $\tilde F_i$ are given in terms
of the BGZV $F_i$ by eq.\sectbp. Again we assume that the BGZV form factors
have been appropriately symmetrised to reflect the symmetries of the basis
operators on which they act.

The BGZV form factors themselves are known algebraic expressions which involve
the following integrals \refs{\BVtwo,\BGZV}:
\eqn\sectde{
f(x) ~~=~~\int_0^1 d\a~e^{-\a(1-\a)x}
}
and
\eqn\sectdf{
F(x_1,x_2,x_3) ~~=~~\int_{\a\ge 0} d^3\a~\d(1-\a_1 -\a_2 -\a_3)~
e^{-(\a_1 \a_2 x_3 + \a_2 \a_3 x_1 + \a_3 \a_1 x_2)}
}

The second-order form factors are easily simplified. Using the BGZV
expressions,
\eqn\sectdg{
f_4(x) = {1\over2}f(x) ~~~~~~~~~~~~~~~~~~f_5(x) = -{1\over 2x}\bigl(f(x)-1\bigr)
}
we have
\eqn\sectdh{
g_0(x) = {1\over 3x} + {1\over x^2} -\Bigl({1\over 2x} + {1\over x^2}\Bigr)f(x) 
}
Using the small $x$ expansion of $f(x)$, viz.
\eqn\sectdi{
f(x) = 1 - {1\over6}x + {1\over60}x^2 -{1\over 840}x^3 + \ldots
}
we find
\eqn\sectdj{
g_0(x) =   {1\over15} - {1\over140}x  + \ldots
}
The absence of inverse powers of $x$ confirms that this is indeed a {\it local}
form factor.

The expansions for the third-order form factors $F_i(x_1,x_2,x_3)$ 
are extremely complicated. They can be read off from the results given 
in ref.\refs{\BGZV} and we will not quote them in full here. Instead, we
focus on a special case of particular interest. In most applications,
it is reasonable to assume that the derivatives of the gravitational 
curvatures are governed by the same scale as the curvatures themselves, 
i.e.~if ${R\over m^2} = O({\l_c^2\over L^2})$ then 
${DR\over m^3} = O({\l_c^3\over L^3})$. In that case, terms with derivatives of 
curvatures are suppressed by the same parameter governing the expansion in
increasing powers of curvature, and should be neglected. We would then only be
interested in the form factors with the first argument set to zero, i.e.
$g_n(0,x_2,x_3)$.

We have calculated the corresponding form factors $F_i(0,x_2,x_3)$ and 
$\tilde F_i(0,x_2,x_3)$ appearing in eq.\sectdd, identifying the new
form factor $\tilde F_0(0,x_2,x_3)$ required for the transformation
to the local basis. The resulting lengthy expressions are collected in 
the appendix.

These results must pass three crucial consistency checks. First, the
transformation from the non-local basis to the local one via the introduction
of $\tilde F_0(x_1,x_2,x_3)$ as described in eq.\sectbp ~must work, i.e.~the
$1/x_1$ singularities in each of the form factors $F_{8,18,19,21}$ must be
removed by the same function $\tilde F_0$. Second, all the resulting form factors
must be entirely local, in the sense that they admit Taylor expansions
with no inverse powers of $x_{1,2,3}$. Finally, the final form factors 
$g_0(0)$ and $g_n(0,0,0),~n=1,2,3$ must reproduce the Schwinger--de Witt
coefficients in the Drummond-Hathrell effective action. In algebraic terms,
each of these tests is highly non-trivial, but as can be verified from the
explicit results in the appendix, each is indeed satisfied.

The $x_{1,2,3}\rta 0$ limit of the form factors in the effective
action can be found from the results quoted in the appendix, using 
$G_n(0,0,0)= -{1\over2}{\a\over\pi}g_n(0,0,0)$. From the definitions 
\sectdd ~we find:
\eqnn\sectdk
$$\eqalignno{
G_0(0) &= -{1\over30}{\a\over\pi} ~~~~~~~~~~
G_1(0,0,0) = -{1\over144}{\a\over\pi}~~~~~~~~ 
G_2(0,0,0) = {13\over360}{\a\over\pi} \cr
G_3(0,0,0) &= -{1\over360}{\a\over\pi}~~~~~~~~~
G_4(0,0,0) = {1\over3360}{\a\over\pi}~~~~~~~~
G_5(0,0,0) = {1\over315}{\a\over\pi} \cr
G_6(0,0,0) &= {1\over2520}{\a\over\pi}~~~~~~~~~~
G_7(0,0,0) = {1\over720}{\a\over\pi}~~~~~~~~~~ 
G_8(0,0,0) = {2\over315}{\a\over\pi}~~~~~~~ \cr
G_9(0,0,0) &= -{5\over1008}{\a\over\pi}   \cr
{}&{}& \sectdk \cr }
$$
and we confirm that $G_0(0)$, $G_{1,2,3}(0,0,0)$ correctly reproduce the 
coefficients $d,a,b,c$ respectively in the Drummond-Hathrell action \sectab.

\vskip0.7cm

\newsec{Numerical Analysis of the Form Factors}

In order to get a better feeling for the behaviour of the form factors,
we have evaluated them numerically. We present some of the results in graphical
form in this section. The general features of the different form factors
are quite similar, so here we just give the analysis for the form factors 
relevant for Ricci flat spaces, i.e. $g_3(0,x_2,x_3)$ and $g_9(0,x_2,x_3)$.

Collecting formulae from the appendix and substituting into eq.\sectdd ~we
find
\eqnn\sectea
$$\eqalignno{
g_3(0,x_2,x_3) ~~=~~&-~F(0,x_2,x_3) {1\over\D}\Bigl[4+\bigl(1+{1\over4}
(x_2+x_3)\bigr)(x_2+x_3)\Bigr] \cr
&+~f(x_2){1\over\D}\Bigl[{3x_2-x_3\over2x_2} + {1\over4}(x_2-x_3)\Bigl(1-
{2(x_2+x_3)^2\over\D}\Bigr) \Bigr] \cr
&+~f(x_3){1\over\D}\Bigl[{3x_3-x_2\over2x_3} + {1\over4}(x_3-x_2)\Bigl(1-
{2(x_2+x_3)^2\over\D}\Bigr) \Bigr] \cr
&+~{1\over\D}\Bigl[-1 + {1\over2}\Bigl({x_3\over x_2} + {x_2\over x_3} + 
x_2 + x_3 \Bigr)\Bigr] ~
+~{1\over12}\Bigl({1\over x_2} + {1\over x_3}\Bigr) ~+~ {1\over4} \Bigl(
{1\over x_2^2} + {1\over x_3^2}\Bigr) \cr
{}&{}& \sectea \cr }
$$
and 
\eqnn\secteb
$$\eqalignno{
g_9(0,x_2,x_3) ~~=~~&-~F(0,x_2,x_3) {1\over\D}\bigl[4 + x_2 + x_3\bigr] \cr
&+~f(x_2)\Bigl[{2\over \D^2}(x_2^2 - x_3^2) - {1\over 2x_2^2} - 
{2\over x_2^3}\Bigr] \cr
&+~f(x_3) \Bigl[-{2\over\D^2}(x_2^2 - x_3^2)\Bigr]~
+~f'(x_2)\Bigl({1\over x_2^2} + {1\over 2x_2}\Bigr) ~-{2\over\D} + 
{1\over 3x_2^2} + {2\over x_2^3} \cr
{}&{}& \secteb \cr }
$$

The behaviour of these functions is illustrated in the following plots
of $g_3(0,x_2,x_3)$ and $g_9(0,x_2,x_3)$. The values at the origin are 
$g_3(0,0,0) = {1\over180}$ and $g_9(0,0,0) = {5\over504}$, as already 
quoted in eq.\sectdk.
Notice that along the diagonals $x_2=x_3$, both functions tend asymptotically 
to zero. This is consistent with the vanishing of the BGZV form factors
as $\square \rta \infty$ as can be seen from the definitions in eqs.\sectde ~and
\sectdf. However, if one argument is set to zero, then 
the functions may tend to a finite limit. It is easy to check that this
behaviour is entirely determined by the $h(x)$ and $h'(x)$ factors in 
eq.\sectdd. These limits are $g_3(0,\infty,0) = g_3(0,0,\infty) = {1\over60}$,
$g_9(0,\infty,0) =0$ and $g_9(0,0,\infty) = {1\over140}$ as can be seen in
the figures below.

\vskip0.6cm
{\epsfxsize=6cm\epsfbox{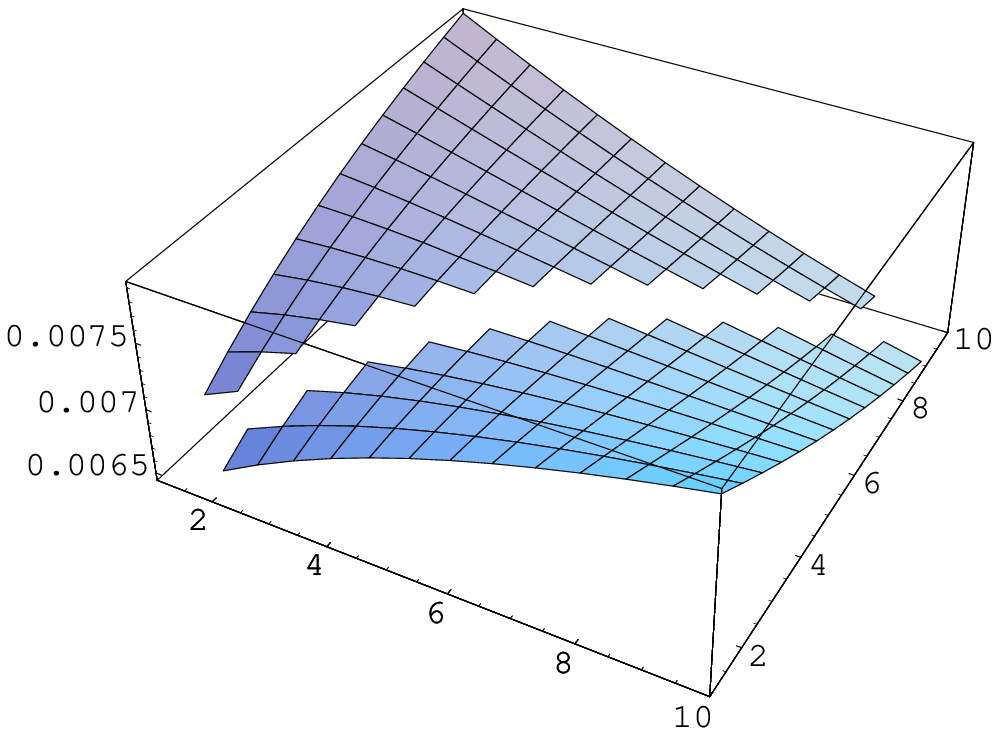}}\hskip1cm
{\epsfxsize=6cm\epsfbox{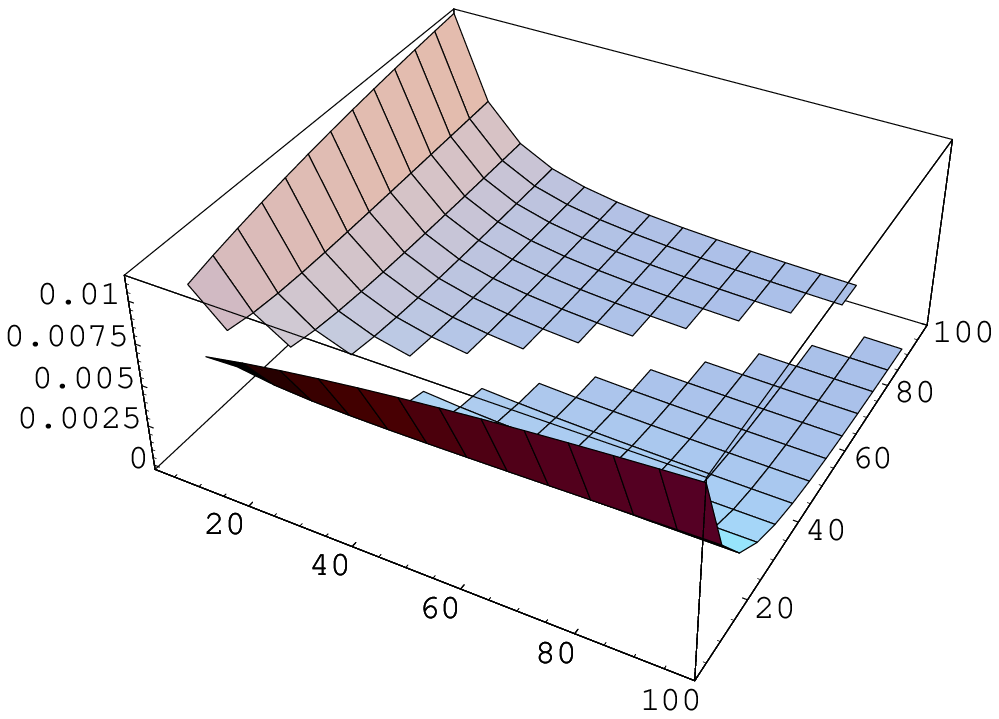}}

\noindent{\eightpoint Fig.2~~3D plots of $g_3(0,x_2,x_3)$ over different ranges.}

\vskip0.6cm
{\epsfxsize=6cm\epsfbox{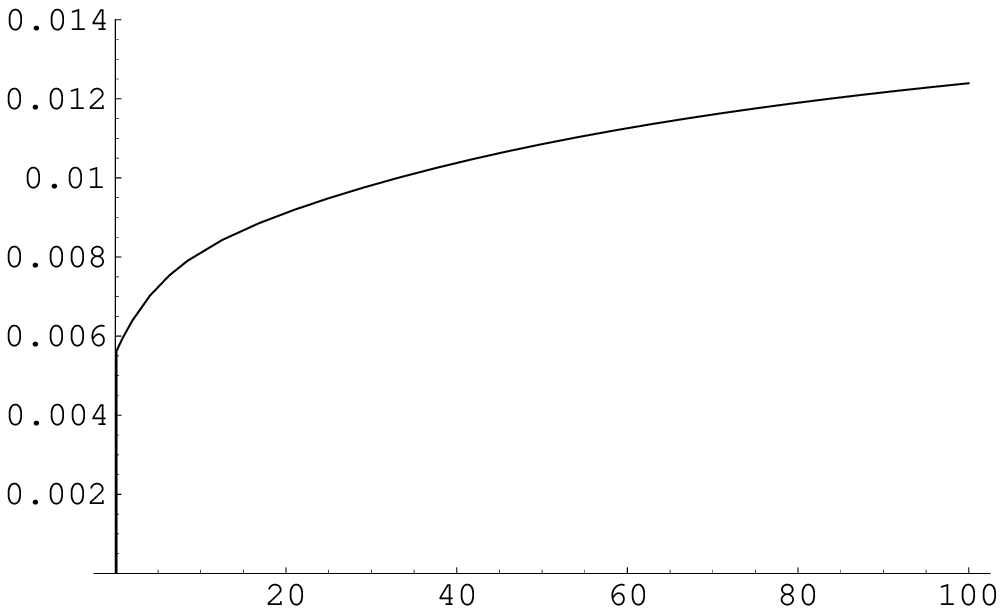}}\hskip1cm
{\epsfxsize=6cm\epsfbox{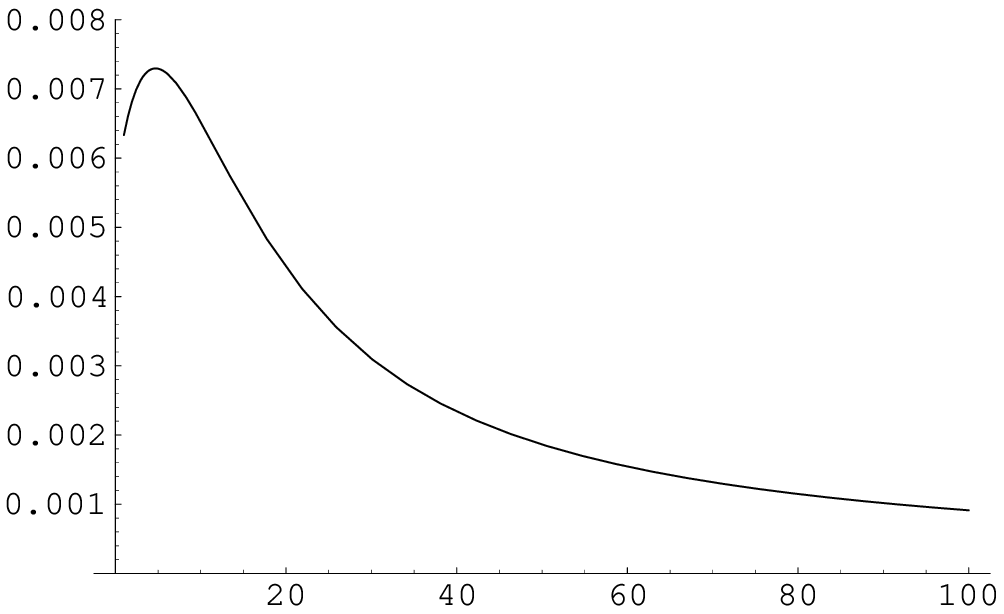}}

\noindent{\eightpoint Fig.3~~Graphs of $g_3(0,x_2,x_3)$ along the $x_2$ or $x_3$ 
axes (left) and the diagonal $x_2=x_3$ (right).}
\vskip0.3cm

\vskip0.6cm
{\epsfxsize=6cm\epsfbox{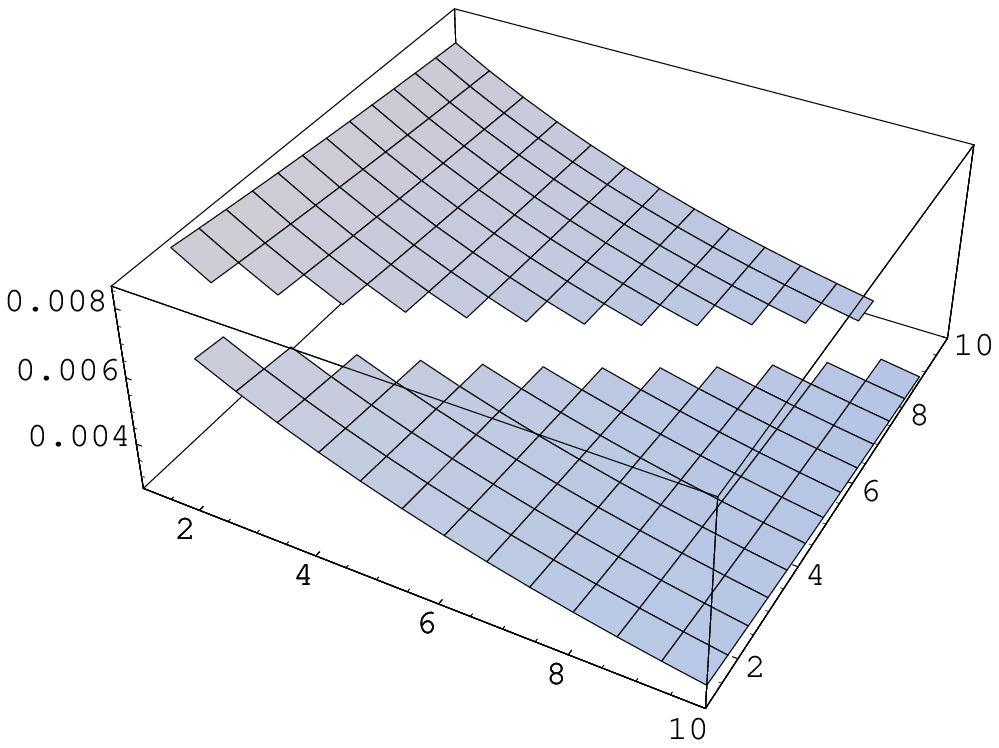}}\hskip1cm
{\epsfxsize=6cm\epsfbox{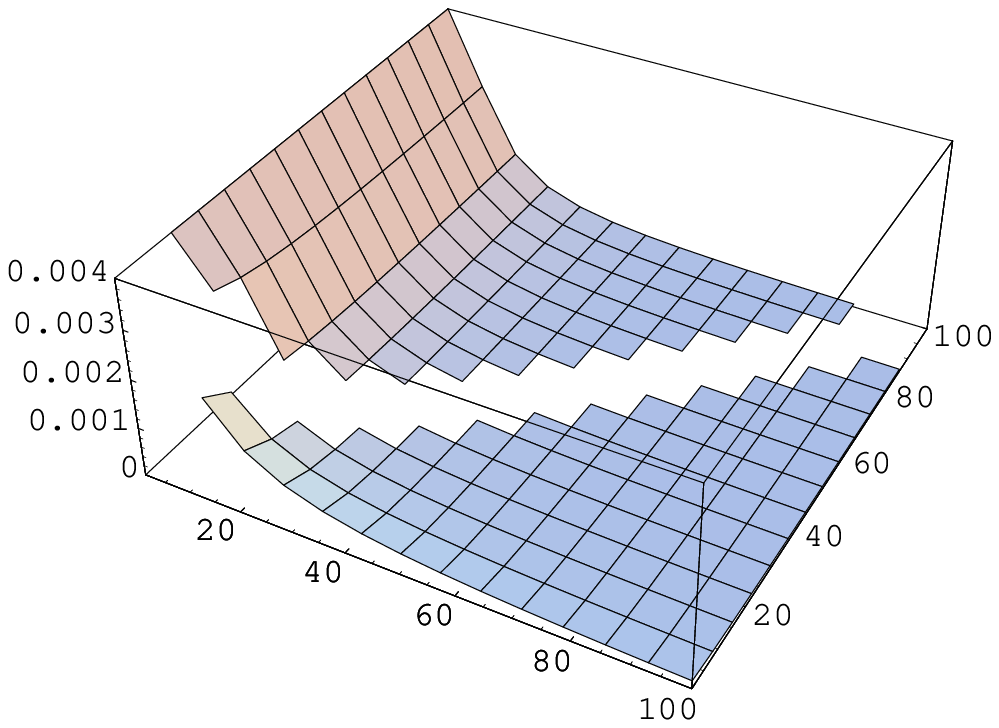}}

\noindent{\eightpoint Fig.4~~3D plots of $g_9(0,x_2,x_3)$ over different ranges.}
\vskip0.3cm

\vskip0.6cm
{\epsfxsize=6cm\epsfbox{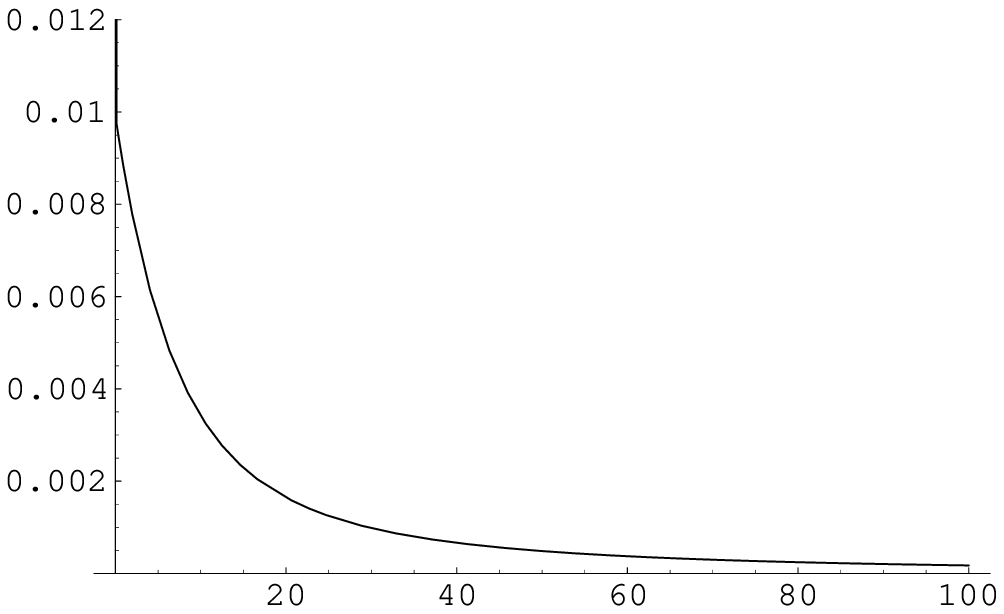}}\hskip1cm
{\epsfxsize=6cm\epsfbox{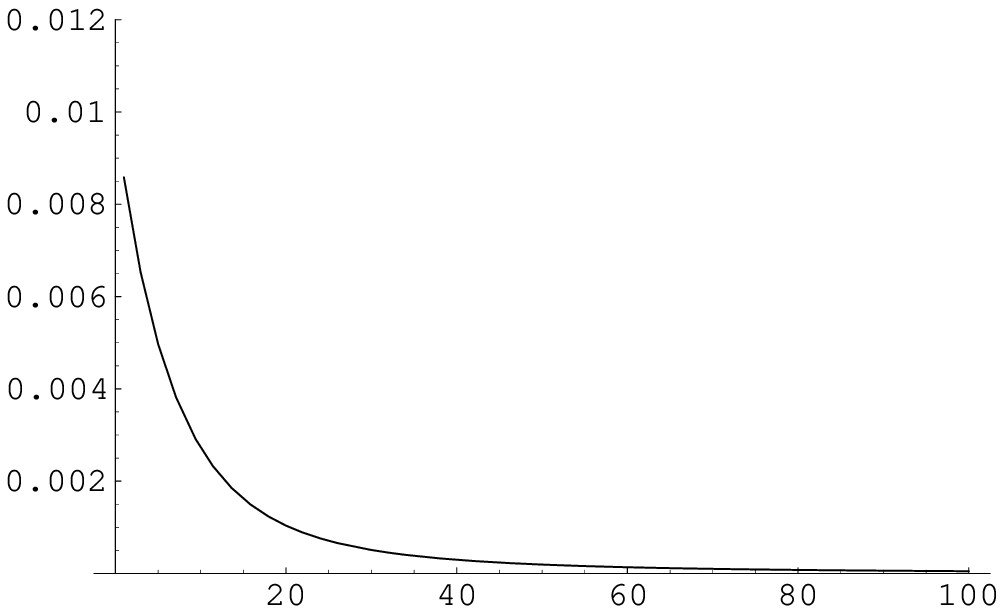}}
\vskip0.6cm
{\epsfxsize=6cm\epsfbox{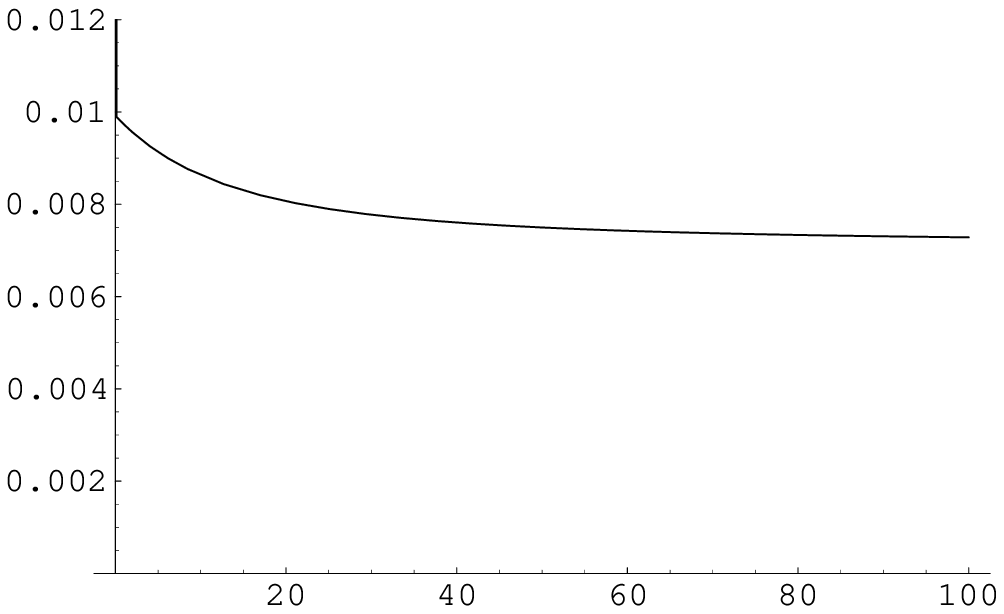}}

\noindent{\eightpoint Fig.5~~Graphs of $g_9(0,x_2,x_3)$ along the $x_2$ 
axis (top left), the diagonal $x_2=x_3$ (top right) and the $x_3$ axis (lower).}
\vskip0.3cm
\vskip0.3cm
This completes our numerical discussion of the form factors. We confirm that
they are regular at $x_i=0$ as required for their role in the local effective
action. Their values at $x_i=0$ match the well-known results for the 
Drummond-Hathrell effective action. They are smooth functions with well-understood
asymptotic behaviour. The local effective action \sectdb ~for photon-gravity 
interactions induced by vacuum polarisation is therefore under complete
algebraic and numerical control to all orders in the derivative expansion. The
precise behaviour of the form factors may now be used to address a variety of 
physical applications, notably the important issue of dispersion for photon
propagation in background gravitational fields \refs{\Ssix}.

\vskip1cm

\noindent{\bf Acknowledgements}

This research is supported in part by PPARC grant 
PPA/G/O/2000/00448.

\vfill\eject

\appendix{A}{}

In this appendix, we quote explicit formulae for the third order form factors 
appearing in the heat kernel in the limit discussed in section 4.

These are expressed in terms of the two basic integrals
\eqn\appxa{
f(x) ~~=~~\int_0^1 d\a~e^{-\a(1-\a)x}
}
and
\eqn\appxb{
F(0,x_2,x_3) ~~=~~{1\over x_2-x_3} ~\int_0^1 d\a~{1\over \a}~
\Bigl( e^{-\a(1-\a)x_3} - e^{-\a(1-\a)x_2} \Bigr) 
}
We will at times use the further limit
\eqnn\appxba
$$\eqalignno{
F(0,x,0) ~~&=~~ {1\over x}~\int_0^1 d\a~{1\over\a}\Bigl(1-
e^{-\a(1-\a)x} \Bigr) \cr
&=~~{1\over2} -{1\over24}x + {1\over360}x^2 -{1\over6720}x^3 + \ldots \cr
{}&{}& \appxba \cr }
$$
The final expressions also involve
\eqn\appxbb{
F'(0,x_2,x_3) ~~\equiv~~{\pl\over\pl x_1} F(x_1,x_2,x_3)\Big|_{x_1=0}
}
which has a small $x$ expansion
\eqn\appxbc{
F'(0,x,0) ~~=~~-{1\over24} + {1\over360}x - {1\over6720}x^2 + \ldots 
}
We also define $\D = (x_2-x_3)^2$. 

It is convenient to recall here the definition of $h(x)$ 
from eqs.\sectcg,\sectdh, viz.
\eqnn\appxbbb
$$\eqalignno{
h(x) ~~&=~~-{1\over6x} -{1\over2x^2} +\Bigl({1\over4x} + 
{1\over2x^2}\Bigr) f(x) \cr
&=~~-{1\over30} + {1\over280}x + \ldots \cr
{}&{}& \appxbbb \cr }
$$

Beginning with the form factors which are not changed in the transformation
to the local basis, we need the following contributions to $g_1(0,x_2,x_3)$.
Note that these have been appropriately symmetrised.
\eqn\appxc{
F_1(0,x_2,x_3) ~~=~~{1\over3} F(0,x_2,x_3)
}
\eqnn\appxd
$$\eqalignno{
F_3(x_2,x_3,0) ~~=~~&F(0,x_2,x_3) \Bigl[-{2\over\D}x_2 x_3 - 
{2\over\D}(x_2+x_3)\Bigr] \cr
&-\Bigl(f(x_2)-f(x_3)\Bigr) {4\over\D^2}x_2 x_3(x_2-x_3) ~+~{1\over\D}(x_2+x_3)\cr
{}&{}& \appxd \cr }
$$
\eqnn\appxe
$$\eqalignno{
F_6(x_2,x_3,0) ~~=~~&F(0,x_2,x_3)\Bigl[ -{1\over6\D}(x_2^2 + 4x_2 x_3 + x_3^2)
-{1\over\D}(x_2+x_3) \Bigr] \cr
&-\Bigl(f(x_2)-f(x_3)\Bigr) {1\over4\D^2}(x_2-x_3) (x_2^2 + 6x_2 x_3 + x_3^2) ~+~
{1\over2\D}(x_2 + x_3) \cr
{}&{}& \appxe \cr }
$$
\eqnn\appxf
$$\eqalignno{
F_7(0,x_2,x_3) ~~=~~&F(0,x_2,x_3)\Bigl[{1\over3\D^2} x_2 x_3(x_2^2+4x_2 x_3+x_3^2)
+{1\over3\D^2}(x_2+x_3)(x_2^2+34x_2 x_3+x_3^2)\Bigr] \cr
&+\Bigl(F(0,x_2,x_3) - {1\over2}\Bigr){2\over\D^2}(x_2^2+10x_2 x_3 + x_3^2)
~-~{1\over\D^2}(x_2+x_3)x_2 x_3 \cr
&+\biggl[{1\over2}\Bigl({f(x_2)-1\over x_2}\Bigr){1\over\D^3}x_2\bigl(
x_2^4 +44x_2^3 x_3 -10 x_2^2 x_3^2 - 36 x_2 x_3^3 + x_3^4\bigr) 
+ x_2\lra x_3\biggr]\cr
&+\biggl[{1\over2}\Bigl({f(x_3)-1\over x_3}\Bigr){1\over\D^3}x_3\bigl(
x_2^4 -36 x_3 x_2^3 -10 x_3^2 x_2^2 + 44 x_3^3 x_2 + x_3^4 \bigr)
+ x_2\lra x_3\biggr]\cr
{}&{}& \appxf \cr }
$$
For $g_4(0,x_2,x_3)$ and $g_9(0,x_2,x_3)$, we require the following:
\eqnn\appxg
$$\eqalignno{
F_{14}(x_2,x_3,0) ~~&=~~F_{14}(0,x_2,x_3) \cr
&=~~F(0,x_2,x_3){2\over\D}(x_2+x_3) ~+~
\Bigl( F(0,x_2,x_3) - {1\over2}\Bigr){8\over\D} \cr
&+~\Bigl(f(x_2) - f(x_3)\Bigr){4\over\D^2}(x_2^2 - x_3^2) \cr
{}&{}& \appxg \cr }
$$
\eqnn\appxh
$$\eqalignno{
F_{20}(0,x_2,x_3) ~~=~~&F(0,x_2,x_3)\Bigl[-{1\over3\D^2}(x_2+x_3)
(x_2^2+4x_2 x_3 +x_3^2)-{2\over3\D^2}(17x_2^2+38x_2 x_3+17x_3^2)\Bigr]\cr
&-\Bigl(F(0,x_2,x_3)-{1\over2}\Bigr){24\over\D^2}(x_2+x_3) ~+~
{1\over\D^2}(x_2+x_3)^2\cr
&-\biggl[f(x_2){2\over3\D^3}(x_2-x_3)(x_2+x_3)(x_2^2+4x_2 x_3+x_3^2) 
+ x_2\lra x_3\biggr]\cr
&-\biggl[\Bigl({f(x_2)-1\over x_2}\Bigr){1\over\D^3}(x_2-x_3)
(21x_2^3 +45x_2^2 x_3 +15x_2 x_3^2 - x_3^3) 
+ x_2\lra x_3\biggr]\cr
{}&{}& \appxh \cr }
$$
For $g_3(0,x_2,x_3)$ we need
\eqn\appxdd{
F_3(0,x_2,x_3) ~~=~~-\Bigl(f(x_2)-f(x_3)\Bigr){1\over2\D}(x_2-x_3)
}
while $g_7(0,x_2,x_3)$ involves
\eqn\appxi{
F_{17}(0,x_2,x_3) ~~=~~-F(0,x_2,x_3) {4\over\D} ~+~
\Bigl(f(x_2) + f(x_3)\Bigr) {1\over\D}
}

We have explicitly checked that these form factors are indeed local
by evaluating the limits $F_i(0,x,0)$ and verifying the cancellation
of all the terms with inverse powers of $x$.

The remaining form factors involve the transformation \sectbp ~between the 
original non-local BGZV form factors and the new local ones. 
The first step is to identify the new form factor $\tilde F_0(0,x_2,x_3)$.
We find this from any of the identities in eqs.\sectbp. For example
(remembering that we understand $F_{18}$ to be symmetrised on $x_2,x_3$
and using the BGZV expression \refs{\BGZV}),
\eqn\appxk{
\tilde F_0(0,x_2,x_3) ~~=~~-{1\over4}~\lim_{x_1\rta 0} ~x_1~
F_{18}(x_1,x_2,x_3) 
}
All the identities produce the same result for $\tilde F_0(0,x_2,x_3)$,
verifying the consistency of our method. We find
\eqnn\appxl
$$\eqalignno{
\tilde F_0(0,x_2,x_3) ~~=~~&- \Bigl( F(0,x_2,x_3) - {1\over2}\Bigr) 
{4\over\D} \cr
&+~ \Bigl(f(x_2) - 1\Bigr) {1\over2\D}{1\over x_2} (3x_2-x_3) ~-~ 
\Bigl(f(x_3) - 1\Bigr) {1\over2\D}{1\over x_3} (x_2-3x_3) \cr
{}&{}& \appxl \cr }
$$
The remaining form factors are then:
\eqnn\appxl
$$\eqalignno{
\tilde F_{18}(0,x_2,x_3) ~~=~~&F(0,x_2,x_3){20\over\D^2} (x_2+x_3)^2 \cr
&+~ \Bigl(F(0,x_2,x_3) - {1\over2}\Bigr){80\over\D^2}(x_2+x_3) ~+~
{16\over\D}F'(0,x_2,x_3) \cr
&+~\biggl[2\Bigl({f(x_2)-1\over x_2}\Bigr){1\over\D^3}(x_2-x_3)
\bigl(17 x_2^3 + 45 x_2^2 x_3 + 19 x_2 x_3^2 - x_3^3\bigr) ~
+ x_2\lra x_3\biggr]\cr
{}&{}& \appxl \cr }
$$
\eqnn\appxm
$$\eqalignno{
\tilde F_{19}(0,x_2,x_3) ~~=~~&-F(0,x_2,x_3){8\over\D^2}(x_2^2+3x_2 x_3+x_3^2)\cr
&-~ \Bigl(F(0,x_2,x_3) - {1\over2}\Bigr){40\over\D^2}(x_2+x_3) ~-~
{8\over\D}F'(0,x_2,x_3) \cr
&+~\biggl[\Bigl({f(x_2)-1\over x_2}\Bigr)\Bigl(
-{8\over\D^3}x_2^2(x_2-x_3)(3x_2+7x_3) \cr
&~~~~~~~~~~+~{12\over\D^4}x_2(x_2+x_3)\bigl(x_2^4-x_2^3 x_3 + 6x_2^2 x_3^2 -
4x_2 x_3^2 +x_3^4\bigr)\Bigr)~
+ x_2\lra x_3\biggr]\cr
{}&{}& \appxm \cr }
$$
\eqnn\appxn
$$\eqalignno{
\tilde F_{21}(0,x_2,x_3) ~~=~~&F(0,x_2,x_3){16\over\D^2}(x_2+x_3)(x_2+4x_3)\cr
&+~ \Bigl(F(0,x_2,x_3) - {1\over2}\Bigr){64\over\D^2}(x_2+4x_3) ~+~
{32\over\D}F'(0,x_2,x_3) \cr
&+~\Bigl({f(x_2)-1\over x_2}\Bigr)\Bigl(
{8\over\D^3}(x_2-x_3)\bigl(9x_2^3+21x_2^2 x_3 +11x_2 x_3^2 -x_3^3\bigr)~-~
{48\over\D^2}x_2(x_2+x_3) \Bigr) \cr
&+~\Bigl({f(x_3)-1\over x_3}\Bigr)\Bigl(
{32\over\D^3}x_3(x_2-x_3)\bigl(x_2^2-6x_2 x_3 -5x_3^2\bigr)~-~
{48\over\D^2}x_3(x_2+x_3) \Bigr) \cr
{}&{}& \appxn \cr }
$$
and finally
\eqnn\appxo
$$\eqalignno{
\tilde F_8(0,x_2,x_3) ~~=~~&-F(0,x_2,x_3){40\over\D^2}x_2 x_3(x_2+x_3)\cr
&-~ \Bigl(F(0,x_2,x_3) - {1\over2}\Bigr){160\over\D^2}x_2 x_3 ~-~
{8\over\D}(x_2+x_3) F'(0,x_2,x_3) \cr
&+~\biggl[\Bigl({f(x_2)-1\over x_2}\Bigr)\Bigl(
-{1\over\D^3}(x_2-x_3)\bigl(17x_2^4 +86x_2^3 x_3 + 72 x_2^2 x_3^2 -22x_2 x_3^3
+7x_3^4\bigr) \cr
&~~~~~~~~~~+~
{6\over\D^4}(x_2+x_3)\bigl(3x_2^2+2x_2 x_3 -x_3^2\bigr) \Bigr)~
+ x_2\lra x_3\biggr]\cr
{}&{}& \appxo \cr }
$$

Again, using the results above, we have explicitly checked that the
form factors $\tilde F_{0,8,18,19,21}$ all have regular small $x$ limits,
as required for the local operator basis. The required numerical cancellations
are complicated and provide a very stringent test of the algebraic
forms quoted.

Their remaining finite values when all of $x_{1,2,3}\rta 0$ reproduce the
coefficients in the Drummond-Hathrell action. To find these, we need the
small $x$ expansions of $f(x)$, $F(0,x,0)$ and $F'(0,x,0)$ given in 
eqs.\sectdi,\appxba,\appxbc. We find:
\eqn\appxp{
\tilde F_0(0,0,0) ~~=~~ \lim_{x\rta 0}~\biggl[-{4\over x^2}
\Bigl(F(0,x,0)-{1\over2}\Bigr) +\Bigl({f(x)-1\over x}\Bigr){3\over2x}
+{1\over12x} \biggr] ~~=~~{1\over72}
}
and similarly
\eqnn\appxq
$$\eqalignno{
F_1(0,0,0) &= {1\over6} ~~~~~~~~~~~~~~F_3(0,0,0) = {1\over12}~~~~~~~~~~~~ 
F_6(0,0,0) = 0 ~~~~~~~~~~~F_7(0,0,0) = 0 \cr
\tilde F_8(0,0,0) &= -{1\over180} ~~~~~~~~
F_{14}(0,0,0) = {1\over180}~~~~~~~~~
F_{17}(0,0,0) = {1\over180} ~~~~~~\tilde F_{18}(0,0,0) = {1\over1260} \cr 
\tilde F_{19}(0,0,0) &= -{1\over1260} ~~~~~~~
F_{20}(0,0,0) = -{1\over7560} ~~~~~
\tilde F_{21}(0,0,0) = {1\over630}\cr
{}&{}& \appxo \cr }
$$

\vfill\eject

\listrefs

\bye